\documentclass[12pt, a4paper]{article}
\usepackage[a4paper, total={6in, 8in}]{geometry}
\usepackage{cite}
\usepackage{amsmath,amssymb,amsfonts}
\usepackage{algorithmic}
\usepackage{graphicx}
\usepackage{textcomp}

\usepackage{cite}
\usepackage{mathtools}
\usepackage{amsfonts} 
\usepackage{booktabs}

\usepackage{authblk}
	\title{Estimation of Dielectric Parameters from Ultrasound Waves in Quantitative Thermoacoustic Tomography}
\author[1]{Teemu Sahlström}
\author[1]{Tanja Tarvainen}
\affil[1]{Department of Technical Physics, University of Eastern Finland, Kuopio, Finland}
\date{}           
\usepackage{amsmath}
\usepackage{amsfonts}
\usepackage{mathtools}
\usepackage{amssymb}

\makeatletter
\def\@maketitle{%
	\newpage\null
	\vspace*{-3em}
	\begin{center}%
		\let \footnote \thanks
		{\LARGE \@title \par}%
		\vskip 1.5em%
		{\large \lineskip .5em%
			\begin{tabular}[t]{c}%
				\@author
			\end{tabular}\par
		}%
		\vskip 1em%
		{\large \@date}%
	\end{center}%
	\par
	\vskip 1.5em}
\makeatother

\begin{document}
	
	\maketitle
	
\begin{abstract}
	Thermoacoustic tomography (TAT) is an imaging technique based on the thermoacoustic effect, combining electromagnetic contrast and high resolution of ultrasound imaging. In TAT, a short micro- or radio wave pulse is directed to the imaged target. Energy of this pulse is absorbed depending on the dielectric parameters of the target, resulting in a spatially varying pressure distribution via the thermoacoustic effect. This pressure, known as the initial pressure distribution, propagates as acoustic waves that are measured on the boundary of the target using ultrasound sensors. In the inverse problem of TAT, the initial pressure is estimated from the measured ultrasound waves. TAT can further be extended to quantitative TAT (QTAT), where the aim is to estimate the dielectric parameters of the target from the measured ultrasound waves, utilizing a model for electromagnetic wave propagation. In this work, we study the inverse problem of QTAT, and propose an approach for simultaneous estimation of electrical conductivity and permittivity from the ultrasound waves. This problem is approached in the framework of Bayesian inverse problems, enabling incorporation of prior and noise models. The forward model describing electromagnetic and acoustic wave propagation is based on the Maxwell's equations and the acoustic wave-equation, respectively. The approach is evaluated with numerical simulations. The results show that the dielectric parameters can be estimated using the proposed approach with good precision. However, the ultrasound sensor geometry and the number of electromagnetic pulses have a significant effect on the accuracy of the estimated parameters.
\end{abstract}

\section{Introduction}
\label{sec:introduction}
Thermoacoustic tomography (TAT) is a biomedical imaging modality based on the thermoacoustic effect, that has been utilized in imaging of soft tissues such as small animals, breast tissues and tumors \cite{lin2021,li2009,fallon2009,patch2016,kellnberger2011,ogunlade2015,li2008,kruger1999}. In TAT, a short micro or radio wave pulse is guided to the imaged target. As the electromagnetic waves propagate through the target, they are absorbed depending on the dielectric properties of the target, leading to thermal expansion and localized increases in pressure. This pressure increase, referred to as the initial pressure distribution, propagates as acoustic waves that are measured on the boundary of the target using ultrasound sensors. A thermoacoustic image, i.e. an estimate of the initial pressure distribution, is then reconstructed from the measured ultrasound waves. TAT therefore combines the electromagnetic contrast arising from interactions of microwaves in of biological tissues with the high resolution of ultrasound imaging.

Thermoacoustic tomography is qualitative by nature and does not allow for quantitative analysis of tissue properties such as various physiological processes. TAT can, however, be extended to its quantitative counterpart (quantitative thermoacoustic tomography, QTAT), where the dielectric parameters of the imaged target are estimated \cite{akhoyari2017,bal2014,sahlstrom2025}. Due to the changes observed in the dielectric parameters in pathologies, such as perfusion changes in ischemic or hemorrhagic stroke \cite{semenov2000, semenov2002}, and various tumors \cite{cheng2018, hussein2019, fernandez2024}, QTAT has significant potential as a diagnostic tool for these, among imaging of other physiological processes. Other techniques for tomographic imaging of dielectric parameters include methods such as microwave tomography (MWT) \cite{semenov2009}, electrical impedance tomography (EIT) \cite{boyle2017}, and magnetic resonance electrical property tomography (MREPT) \cite{zhang2014}. Due to the coupled physics of electromagnetic and ultrasound wave propagation, QTAT has the potential to estimate the dielectric parameters with a higher resolution compared to MWT and EIT, and with a lower financial cost compared to MREPT.

The inverse problem of QTAT can be approached in two stages. First, the initial pressure distribution is estimated from the measured ultrasound waves by solving the acoustic inverse problem. Then, in the electrical inverse problem, the dielectric parameters are estimated from the initial pressure distribution. The acoustic inverse problem, that is similar to to the inverse problem in photoacoustic tomography, has been widely studied, and various approaches for its solution have been considered. These approaches can be broadly categorized in analytical approaches, approaches utilizing a numerical approximation of the forward model such as time-reversal, least-squares approaches and the Bayesian approach, and deep learning based techniques, see e.g. \cite{poudel2019survey,kuchment2011bookchapter,hauptmann2020,tick2016} and the references therein. 

The previous research on the electrical inverse problem of QTAT has mostly focused on its theoretical foundations, and it has not been realized in practice. Uniqueness and stability results for estimating the conductivity in the electrical inverse problem using the vectorial Maxwell's equation and the Helmholtz equation were given in \cite{bal2011} and a reconstruction approach using the Helmholtz equation was presented in \cite{ammari2013}. Furthermore, uniqueness results for estimating both conductivity and permittivity in the electrical inverse problem for the vectorial Maxwell's equation were given in \cite{bal2014} and uniqueness and stability results for a system of semilinear Helmholtz equations were presented in \cite{ren2023}. Notably, in \cite{bal2014} it was shown that multiple incident electric fields inducing sufficiently different initial pressure distribution are required for obtaining a unique solution to the electrical inverse problem. For numerical implementations, results for estimating the conductivity using the Helmholtz and Maxwell's equations were presented in \cite{bal2011}, and estimating the Gru\"neisen parameter, susceptibility and conductivity using the system of semilinear Helmholtz equations were presented in \cite{ren2023}. Finally, numerical results for estimating the conductivity and permittivity using the Maxwell's equations were presented in \cite{sahlstrom2025}.

Only few studies have considered the full inverse problem of QTAT of estimating the dielectric parameters from ultrasound waves. A model taking into account slowly varying electromagnetic sources and a formal approach for the inverse problem was presented in \cite{akhoyari2017}, and the uniqueness and stability for reconstructing small conductivity inclusions was studied in \cite{jebawy2020}. 

In this work, we propose an approach for estimation of electrical conductivity and permittivity from the ultrasound waves in the full two-stage inverse problem of QTAT, building on our previous work on the electrical part \cite{sahlstrom2025}. The problem is approached in the framework of Bayesian inverse problems, enabling formulation of the prior model for the acoustic inverse problem, and modelling of noise propagation from data and solution of the acoustic inverse problem to the electrical inverse problem. The forward models are based on the Maxwell's equations and the acoustic wave equation, that are numerically approximated using the finite element method (FEM) and the pseudospectral $k$-space method, respectively. The approach is evaluated with numerical simulations using a varying number of electromagnetic sources, and full and limited view ultrasound sensor geometries.

The remainder of the paper is organized as follows. Modeling of electromagnetic and acoustic wave propagation are presented in Sec.\ref{sec:Forward problem}. The acoustic and electrical inverse problems are described in Sec. \ref{sec:Inverse problem} and numerical simulations are presented in Sec. \ref{sec:Simulations}. The results are discussed and conclusions are drawn in Sec. \ref{sec:Discussion}.

\section{Forward problem}
\label{sec:Forward problem}
In the forward problem of QTAT, the ultrasound waves on the boundary of the target are solved when the dielectric parameters are given. The forward problem is approached in two stages known as the electrical and acoustic forward problems.

\subsection{Electrical forward problem}
\label{subsec:Electromagnetic forward model}
In the electrical forward problem, the initial pressure distribution is solved when the dielectric parameters of the target and the electromagnetic source are given. Let us consider a domain $\Omega_\mathrm{el} \subset \mathbb{R}^d$ with a boundary $\partial \Omega$, where $d$ denotes the dimension. We assume inhomogeneous, non-magnetic, linear and isotropic materials with conductivity $\sigma(r) \geq 0$ and relative permittivity $\epsilon(r) \geq 1$. We further model the electric fields as time-harmonic \cite{book_monk}
\begin{equation}
	\hat{E}(r,t) = \mathrm{Re}(E(r)e^{\mathrm{i}\omega t}),
\end{equation}
where $\hat{E}(r,t)$ is the electric field at time instance $t$ and position $r$, $\mathrm{Re}(\cdot)$ denotes the real part, $E(r) \in \mathbb{C}^d$ is the complex amplitude, $\mathrm{i}$ is the imaginary unit, and $\omega$ is the angular frequency. 

Propagation of electromagnetic waves is governed by the Maxwell's equations. In this work, we assume that the exterior of $\Omega_\mathrm{el}$ has constant electrical conductivity $\sigma_\mathrm{ext} \geq 0$ and relative permittivity $\epsilon_{r,\mathrm{ext}} \geq 1$ and use an absorbing type boundary condition. The behavior of electric fields is then described by the system of equations \cite{book_monk, thesis_bonazolli}
\begin{align}
	\begin{dcases}
		\nabla \times \nabla \times E(r) - \gamma^2(r) E(r) = 0, \quad r \in \Omega_\mathrm{el}\\
		(\nabla \times E(r)) \times \hat{n} + \mathrm{i}\kappa \hat{n} \times (E(r) \times \hat{n}) = g(r),  \quad r \in \partial \Omega_\mathrm{el},
	\end{dcases}
	\label{eq:double_curl}
\end{align}
where $\gamma^2(r) = \mu_0 ( \omega^2\epsilon_0\epsilon_r(r) - \mathrm{i} \omega \sigma(r) )$, $\epsilon_0$ is the permittivity of free space, $\mu_0$ is the permeability of free space, and $\epsilon_r(r)$ is the relative permittivity. Furthermore, $\kappa = \sqrt{\mu_0 ( \omega^2\epsilon_0\epsilon_{r,{\rm ext}} - \mathrm{i} \omega \sigma_{\rm ext} )}$, and $\hat{n}$ is an outward unit normal. The right hand side term of the boundary condition describing the electromagnetic source is given by $g(r) = (\nabla \times E_{\mathrm{inc}}(r)) \times \hat{n} + \mathrm{i}\kappa \hat{n} \times (E_{\mathrm{inc}}(r) \times \hat{n})$, where $E_\mathrm{inc}$ is the incident electric field on the boundary $\partial \Omega_\mathrm{el}$.

As the electric fields propagate through the domain, they are absorbed leading to an absorbed energy density \cite{li2009,akhoyari2017}
\begin{equation}
	H(r,t) = \sigma(r) \langle \vert \hat{E}(r,t) \vert^2 \rangle,
\end{equation}
where $\langle \cdot \rangle$ denotes a short time average over one oscillatory period $T = 2\pi / \omega$ of the electric field. When the absorption of electromagnetic energy is assumed as instantaneous, i.e. it fulfills the so-called thermal and stress confinement criteria \cite{photoacousticimaging_book}, the absorbed energy density can be written as \cite{akhoyari2017,ren2023,bal2014}
\begin{equation}
	H(r) = \frac{1}{2} \sigma(r)\vert E(r) \vert ^2.
	\label{eq:absorbed_energy_density}
\end{equation}
The initial pressure $p_0$ is then given in terms of the absorbed energy density as \cite{li2009,photoacousticimaging_book}
\begin{equation}
	p_0(r) = \hat{\Gamma} H(r),
	\label{eq:initial_pressure}
\end{equation}
where $\hat{\Gamma}$ is the Gr\"uneisen parameter describing the efficiency of the thermoacoustic effect. In this work, the Gr\"uneisen parameter is assumed to be a known constant.

In this work, the solution of \eqref{eq:double_curl} is approximated using the FEM with edge elements \cite{book_monk,sahlstrom2025,thesis_bonazolli}. Edge elements are especially suitable for modelling the behavior of electromagnetic fields in complex media due to their ability of incorporating continuity properties and avoidance of spurious numerical solutions. The FEM is implemented in MATLAB utilizing a code package published in \cite{chen2008}.

\subsection{Acoustic forward problem}
\label{subsec:Acoustic forward model}
In the acoustic forward problem, the ultrasound waves are solved from the initial pressure distribution given by the solution of the electrical forward problem. Propagation of acoustic waves generated by an initial pressure distribution in a domain $\Omega_\mathrm{ac} \subset \mathbb{R}^d$ in an acoustically homogeneous and non-attenuating medium is described by the acoustic initial value problem \cite{cox2005,treeby2010}
\begin{equation}
	\begin{dcases}
		\nabla^2p(r,t) - \frac{1}{v^2}\frac{\partial^2 p(r,t)}{\partial t^2} = 0 \\
		p(r,t=0) = p_0(r) \\
		\frac{\partial}{\partial t}p(r,t = 0) = 0,
	\end{dcases}
	\label{eq:acoustic_initia_value_problem}
\end{equation}
where $p(r,t)$ is the pressure, $p_0(r)$ is the initial pressure, and $v$ is the speed of sound. In this work, the solution of \eqref{eq:acoustic_initia_value_problem} is numerically approximated using the pseudospectral $k$-space method implemented in the k-Wave toolbox \cite{treeby2010}.

\section{Inverse problem}
\label{sec:Inverse problem}
In the inverse problem of QTAT, the dielectric parameters of the imaged target are estimated from the measured ultrasound waves. The problem is approached in two stages by solving the acoustic and electrical inverse problems. 

\subsection{Acoustic inverse problem}
\label{subsec:Acoustic inverse problem}
In the acoustic inverse problem of QTAT, the initial pressure distribution is estimated from the measured ultrasound waves. Let us consider a thermoacoustic measurement system consisting of $S$ ultrasound sensors measuring an ultrasound signal discretized in $T$ time points. The discretized observation model for the acoustic inverse problem is then given by 
\begin{equation}
	p_t = Kp_0 + e,
	\label{eq:acoustic_observation_model}
\end{equation}
where $p_t \in \mathbb{R}^{ST}$ is the ultrasound data, $K \in \mathbb{R}^{ST \times L}$ is a discretized linear forward operator describing the acoustic wave equation \eqref{eq:acoustic_initia_value_problem}, $p_0 \in \mathbb{R}^{L}$ is the discretized initial pressure distribution, and $e \in \mathbb{R}^{ST}$ denotes additive noise. In this work, the forward operator $K$ is constructed by simulating impulse responses for the $L$ discretization points (pixels) in the domain using the pseudospectral $k$-space method implemented with the k-Wave toolbox and placing the resulting simulated waveforms on the columns of the matrix $K$ \cite{tick2016,sahlstrom2021}.

The inverse problem is approached in the Bayesian framework. The solution to the inverse problem is the posterior distribution $\pi(p_0 \vert p_t)$, given by the Bayes' formula \cite{kaipio_book,book_calvetti}
\begin{equation}
	\pi(p_0 \vert p_t) \propto \pi(p_t \vert p_0) \pi(p_0),
	\label{eq:posterior_distribution}
\end{equation}
where $\pi(p_t \vert p_0)$ is the likelihood distribution and $\pi(p_0)$ is the prior distribution of the initial pressure. In the case of mutually independent noise and initial pressure, the likelihood distribution can be written as \cite{kaipio_book}
\begin{equation}
	\pi(p_t \vert p_0) = \pi_e(p_t - Kp_0),
	\label{eq:acoustic_likelihood}
\end{equation}
where $\pi_e$ denotes the probability density of the noise. We model the noise and the initial pressure as Gaussian distributed
\begin{equation*}
	e \sim \mathcal{N}(\eta_{e}, \Gamma_{e}), \quad p_0 \sim \mathcal{N}(\eta_{p_0}, \Gamma_{p_0}),
	\label{eq:acoustic_gaussian_assumptions}
\end{equation*}
where $\eta_{e}$ and $\eta_{p_0}$ are the expected values, and $\Gamma_{e}$ and $\Gamma_{p_0}$ are the covariance matrices of the noise and initial pressure distributions, respectively. 

In the two-stage inverse problem considered in this work, the prior information for the first (acoustic) inverse problem, can be determined by the prior model of the dielectric parameters and the electrical forward model. In this work, the prior model for the acoustic inverse problem is formed for each incident electric field as a Gaussian sample-based approximation \cite{kaipio_book}
\begin{align}
	\eta_{p_0} &= \frac{1}{N} \sum_{n=1}^N p_{0,n}, 
	\label{eq:sample_prior_eta} \\ 
	\Gamma_{p_0} &= \frac{1}{N} \sum_{n=1}^N p_{0,n}p_{0,n}^\mathrm{T} - \eta_{p_0}\eta_{p_{0}}^{\mathrm{T}},
	\label{eq:sample_prior_gamma}
\end{align}
where $p_{0,n}$ are samples of the initial pressure distribution calculated using the electrical forward model (Eqs. \eqref{eq:double_curl}, \eqref{eq:absorbed_energy_density}-\eqref{eq:initial_pressure}) from samples drawn from the prior distributions of the dielectric parameters described in Sec. \ref{subsec:Electrical inverse problem theory}.  

The posterior distribution can then be written as\cite{kaipio_book,tick2016}
\begin{equation}
	\pi(p_0 \vert p_t) \propto \exp \bigg\{ -\frac{1}{2} \Vert L_e(p_t - Kp_0 -\eta_e) \Vert _2^2 
	-\frac{1}{2} \Vert L_{p_0}(p_0 - \eta_{p_0}) \Vert _2^2  \bigg\},
	\label{eq:acoustic_posterior}
\end{equation}
where $L_{e}$ and $L_{p_0}$ are the Cholesky decompositions of the inverse covariance matrices of the noise and prior distributions $\Gamma_e^{-1} = L_e^\mathrm{T}L_e$ and $\Gamma_{p_0}^{-1} = L_{p_0}^\mathrm{T}L_{p_0}$. Due to the linearity of the observation model and Gaussian distributed noise and prior, the posterior distribution is a Gaussian distribution
\begin{equation}
	p_0 \vert p_t \sim \mathcal{N}(\eta_{p_0 \vert p_t}, \Gamma_{p_0 \vert p_t}),
\end{equation}
where the expected value $\eta_{p_0 \vert p_t}$ and covariance matrix $\Gamma_{p_0 \vert p_t}$ are given by
\begin{align}
	\eta_{p_0 \vert p_t} &= \Gamma_{p_0 \vert p_t} (K^\mathrm{T} \Gamma_{e}^{-1} (p_t-\eta_{e}) + \Gamma_{p_0}^{-1}\eta_{p_0}),
	\label{eq:acoustic_posterior_mean}\\
	\Gamma_{p_0 \vert p_t} &= (\Gamma_{p_0}^{-1} + K^\mathrm{T}\Gamma_{e}^{-1}K)^{-1}.
	\label{eq:acoustic_posterior_cov}
\end{align}

\subsection{Electrical inverse problem}
\label{subsec:Electrical inverse problem theory}
In the electrical inverse problem, the dielectric parameters are estimated from the initial pressure distributions that are obtained as the solutions of the acoustic inverse problem. Let us consider initial pressure data induced by $M$ incident electric fields $P_0 = (p_{0,1}, \dots, p_{0,M})^\mathrm{T} \in \mathbb{R}^{ML}$
where $p_{0,m}= \eta_{p_0 \vert p_t,m}$ is the expected value of the posterior distribution of the initial pressure corresponding the $m$:th incident electric field given by \eqref{eq:acoustic_posterior_mean}. The discretized observation model for the electrical inverse problem is of the form \cite{sahlstrom2025}
\begin{equation}
	P_0 = f(x) + \tilde{e},
	\label{eq:electrical_observation_model}
\end{equation}
where $x = (\sigma_1, \dots, \sigma_J, \epsilon_{r,1}, \dots, \epsilon_{r,J})^\mathrm{T} \in \mathbb{R}^{2J}$ are the discretized dielectric parameters, $f$ is a discretized forward operator, Eqs. \eqref{eq:double_curl}, \eqref{eq:absorbed_energy_density}-\eqref{eq:initial_pressure}, and $\tilde{e} \in \mathbb{R}^{ML}$ is additive noise. 

Similarly as in the acoustic inverse problem, the electrical inverse problem is approached in the Bayesian framework. The posterior distribution is given by  \cite{kaipio_book}
\begin{equation}
	\pi(x \vert P_0) \propto \pi(P_0 \vert x) \pi(x),
\end{equation}
where $\pi(P_0 \vert x)$ is the likelihood distribution and $\pi(x)$ is the prior distribution of the dielectric parameters.

The noise and unknown parameters are modelled as Gaussian distributed and mutually independent
\begin{equation*}
	\tilde{e} \sim \mathcal{N}(\eta_{\tilde{e}}, \Gamma_{\tilde{e}}), \quad x \sim \mathcal{N}(\eta_{x}, \Gamma_{x}),
	\label{eq:electrical_gaussian_assumptions}
\end{equation*}
where $\eta_{\tilde{e}}$ and $\eta_x$ are the expected values, and $\Gamma_{\tilde{e}}$ and $\Gamma_x$ are the covariance matrices of the noise and dielectric parameter distributions, respectively. Now, in the case of the two-stage inverse problem, the noise is modeled as zero mean, with a covariance matrix given by the solution of the acoustic inverse problem
\begin{equation}
	\eta_{\tilde{e}} = 0, \quad \Gamma_{\tilde{e}} = \mathrm{diag}( \Gamma_{p_0 \vert p_t,1}, \dots  \Gamma_{p_0 \vert p_t,M}),
	\label{eq:electrical_noise_statistics}
\end{equation}
where $\Gamma_{p_0 \vert p_t,m}$ is the covariance matrix of the initial pressure distribution given by \eqref{eq:acoustic_posterior_cov} corresponding to the $m$:th incident electric field, and $\mathrm{diag}(\cdot)$ denotes a block diagonal matrix. The prior model for the dielectric parameters used in this work is the Gaussian Ornstein-Uhlenbeck distribution \cite{rasmussen_book} with a covariance
\begin{equation}
	\Gamma_x =
	\begin{bmatrix}
		\tilde{\sigma}^2_\sigma \Pi & 0 \\ 0 & \tilde{\sigma}^2_{\epsilon_r} \Pi,
		\label{eq:dielectric_prior_Gamma}
	\end{bmatrix},
\end{equation}
where $\tilde{\sigma}_\sigma$ and $\tilde{\sigma}_{\epsilon_r}$ are the standard deviations of the conductivity and relative permittivity, respectively. Furthermore, $\Pi$ is defined as
\begin{equation}
	\Pi = \exp \left\{ - \frac{\Vert r_i - r_j \Vert}{\ell} \right\},
	\label{eq:dielectric_prior_Pi}
\end{equation}
where $\ell$ is a characteristic length describing spatial correlation, and $r_i$ and $r_j$ are positions of discretization points (element centers).

The posterior distribution can then be written as \cite{kaipio_book,sahlstrom2025}
\begin{equation}
	\pi(x \vert P_0) \propto \exp \bigg\{ -\frac{1}{2} \Vert L_{\tilde{e}}(P_0 - f(x)) \Vert _2^2 -\frac{1}{2} \Vert L_{x}(x - \eta_{x}) \Vert _2^2  \bigg\},
	\label{eq:electrical_posterior}
\end{equation}
where $L_{x}$ and $L_{\tilde{e}}$ are the Cholesky decompositions of the inverse covariance matrices of the prior and noise distributions $\Gamma_{x}^{-1} = L_{x}^\mathrm{T}L_{x}$ and $\Gamma_{\tilde{e}}^{-1} = L_{\tilde{e}}^\mathrm{T}L_{\tilde{e}}$, respectively. The posterior distribution \eqref{eq:electrical_posterior} could, in principle, be evaluated using Markov-chain Monte Carlo methods. These methods are, however, computationally prohibitive in high dimensional inverse problems. We therefore consider point estimates and compute the \emph{maximum a posteriori} (MAP) estimate
\begin{equation}
	x_{\mathrm{MAP}} = \arg\min_x \bigg\{ \frac{1}{2} \Vert L_{\tilde{e}}(P_0 - f(x)) \Vert _2^2 
	+\frac{1}{2} \Vert L_{x}(x - \eta_{x}) \Vert _2^2  \bigg\}.
	\label{eq:map_estimate}
\end{equation}
In this work, the MAP estimate is solved using the Gauss-Newton method, where the solution is obtained iteratively as
\begin{equation}
	x_{l+1} = x_l + \tilde{\alpha}_ld_l,
	\label{eq:gauss_newton_iteration}
\end{equation}
where $\tilde{\alpha}_l$ is a step length on the $l$:th iteration. Furthermore, $d_l$ is a search direction given by
\begin{equation}
	d_l = \left(  J_{f(x_l)}^\mathrm{T} \Gamma_{\tilde{e}}^{-1} J_{f(x_l)} + \Gamma_x^{-1}  \right)^{-1}
	\left( J_{f(x_l)}^\mathrm{T} \Gamma_{\tilde{e}}^{-1} (H-f(x_l)) - \Gamma_x^{-1}(x - \eta_x) \right),
\end{equation}
where $J_{f(x_l)} \in \mathbb{R}^{ML \times 2J}$ is the Jacobian matrix of the forward operator $f$ evaluated at the point $x_l$.

\section{Simulations}
\label{sec:Simulations}
\subsection{Simulation domain}
\label{subsec:Simulation domain}
\begin{figure}[!tbp]
	\centering
	\includegraphics[width=0.6\linewidth]{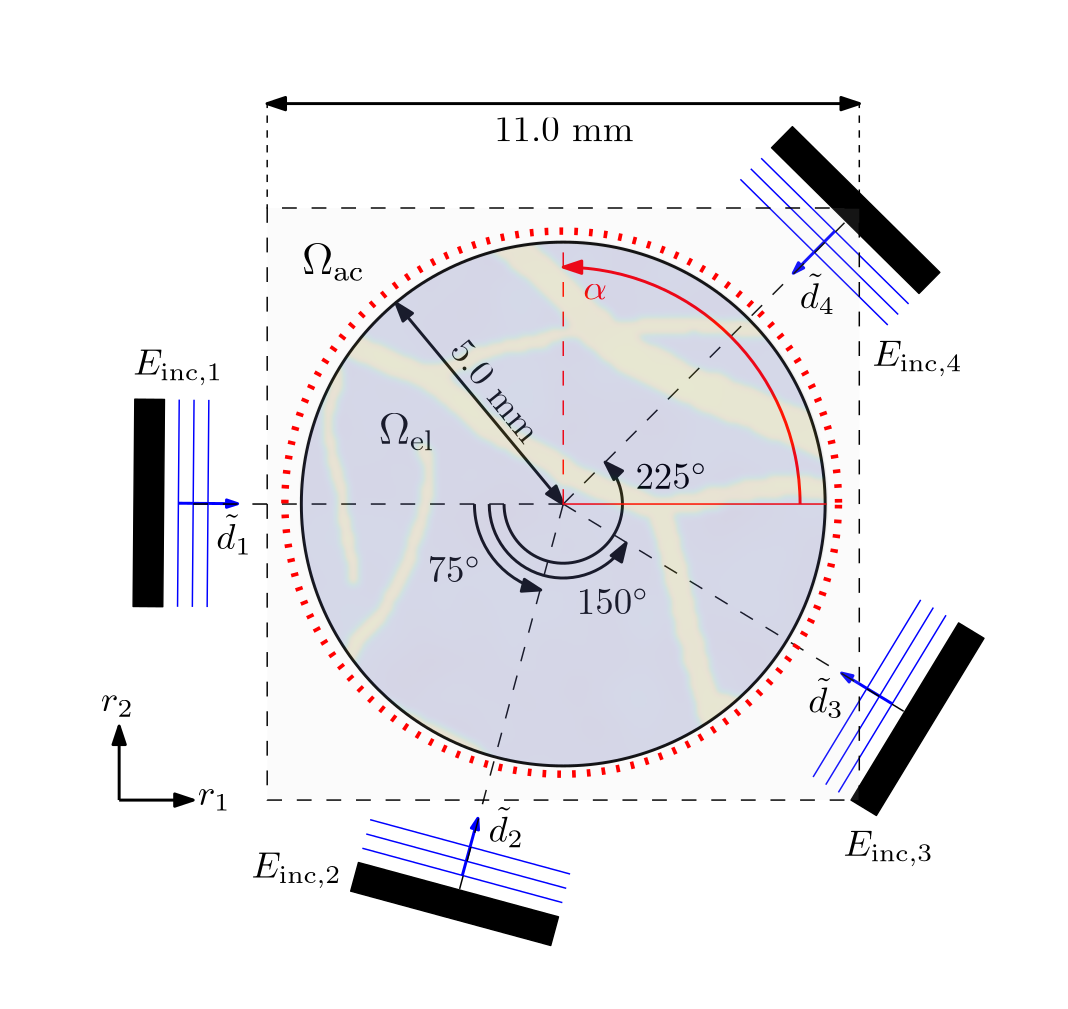}
	\caption{An illustration of the simulation setup. The simulated phantom (circle), and electromagnetic and acoustic domains $\Omega_\mathrm{ac}$ and $\Omega_\mathrm{el}$. The incident electric fields $E_{\mathrm{inc},m}$ are illustrated using black rectangles and propagation directions $d_m$, where $m = 1, \dots, 4$. The ultrasound sensors geometries are shown using red dots and their angular coverage is indicated using the angle $\alpha$.}
	\label{fig:setup}
\end{figure}

The proposed approach for estimating the dielectric parameters from ultrasound waves in QTAT was studied using 2D numerical simulations. The simulated numerical phantom was defined as a circle with a radius of 5mm. The incident electric fields were modeled as linearly polarized plane waves from four directions around the phantom. Ultrasound sensors were modeled as point-like sensors located on a circular arc at a distance of 0.125 mm from the phantom boundary. The simulated phantom, incident electric fields, and ultrasound sensor geometries are illustrated in Fig. \ref{fig:setup}. The electromagnetic simulation domain $\Omega_{\mathrm{el}}$ was modeled as the size of the phantom. For ultrasound simulations, a 11.0 $\times$ 11.0 mm square domain $\Omega_{\mathrm{ac}}$ centered at the phantom location was considered. The acoustic domain was further extended using a perfectly matched layer (PML) to absorb reflections on the boundary $\partial \Omega_{\mathrm{ac}}$.

\subsection{Data simulation}
\label{subsec:Data simulation}
\begin{figure}[!tbp]
	\centering
	\includegraphics{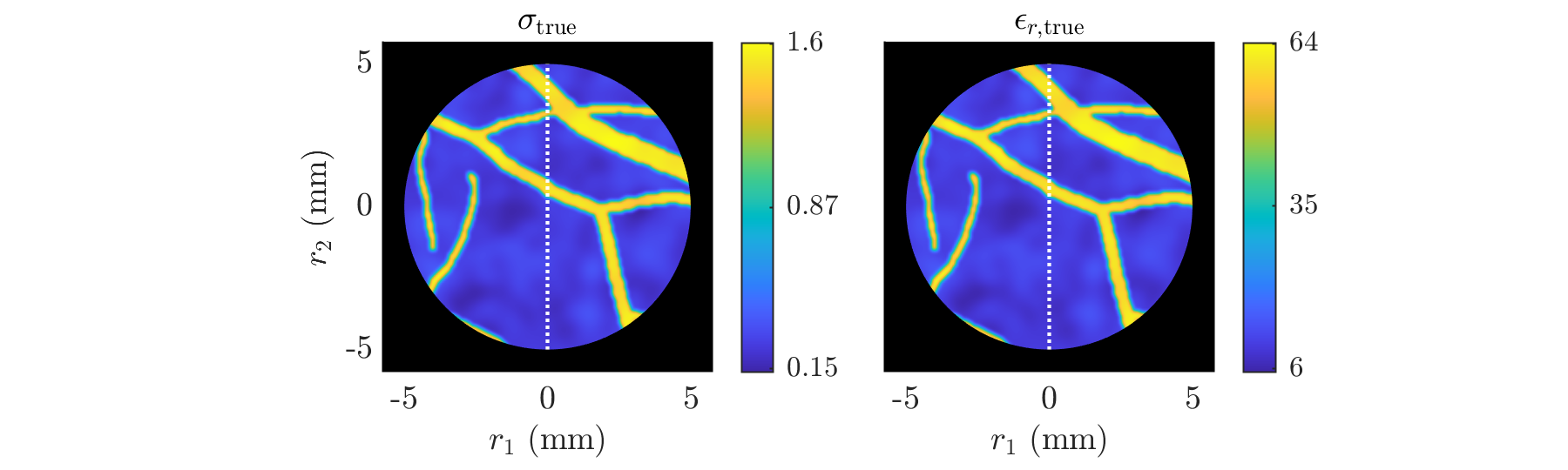}
	\caption{Simulated conductivity $\sigma_{\mathrm{true}}$ (Sm$^{-1}$, left) and relative permittivity $\epsilon_{r,\mathrm{true}}$ (right). Location of the cross section where the results are visualised is illustrated with a dotted white line.}
	\label{fig:phantoms}
\end{figure}

Thermoacoustic ultrasound data was simulated using a numerical phantom consisting of blood vessel like structures shown in Fig.\ref{fig:phantoms}. Values for the dielectric parameters were chosen in the range of realistic values for biological tissues \cite{gabriel1996}. The minimum and maximum values of the dielectric parameters were $\sigma_{\mathrm{min}} = 0.14\,\mathrm{Sm^{-1}}\,,\sigma_{\mathrm{max}} = 1.60\,\mathrm{Sm^{-1}}\,, \epsilon_{r,\mathrm{min}} = 5.44,$ and $\epsilon_{r,\mathrm{max}} = 64.03$. Exterior of the domain was assumed to consists of a food grade oil, such as castor oil, with dielectric parameters $\sigma_\mathrm{ext} = 10^{-12} \, \mathrm{Sm}^{-1}$ and $\epsilon_{r, \mathrm{ext}} = 4$.

For the electromagnetic simulations, $M = 4$ incident electric fields were considered. The incident electric fields were modeled as linearly polarized plane waves of the form
\begin{equation}
	E_{\mathrm{inc},m}(r) = \tilde{p}_m \exp \left\{ -\mathrm{i}\kappa \tilde{d}_m\cdot r \right\},
\end{equation}
where $m = 1, \dots 4$, $\tilde{p}$ is the polarisation, $\tilde{d} \in \mathbb{R}^2$ is the propagation direction, and $\Vert \tilde{d} \Vert = 1$. The plane waves were polarized perpendicularly to the propagation direction, i.e. $\tilde{d}_m\cdot \tilde{p}_m = 0$. The incident electric field directions are illustrated in Fig. \ref{fig:setup}. The initial pressures were simulated using the Maxwell's equation \eqref{eq:double_curl} and the model \eqref{eq:absorbed_energy_density}--\eqref{eq:initial_pressure} using the FEM as described in Sec. \ref{subsec:Electromagnetic forward model} in a triangular FE discretization described in Table \ref{tab:discretisations_electromagnetic}.
The Gr\"uneisen parameters was $\hat{\Gamma}=1$ and the frequency was $f=$ 1 GHz. The simulated initial pressure distributions are illustrated later in Fig. \ref{fig:acoustic}.
\begin{table}[!tbp]
	\caption{Spatial discretisation used for data simulation and the electrical inverse problem. Number of elements $J$, number of edges $N_e$, number of nodes $N_n$, and average edge length $\Delta h$.}
	\label{tab:discretisations_electromagnetic}
	\centering	
	\begin{tabular}{lcccc}		
		Electrical & $J$ & $N_e$ & $N_n$ & $\Delta h\, (\mu\mathrm{m})$ \\ 
		\midrule
		Simulation & 536215 & 805239 & 269025 & 19.5  \\
		Inverse    & 19555 & 29507 & 9953 & 101.9 \\
		\midrule
	\end{tabular}
\end{table}

The initial pressures were then linearly interpolated to a pixel mesh described in Table \ref{tab:discretisations_acoustic} using the inbuilt \texttt{scatteredInterpolant} MATLAB function. The ultrasound data was simulated with the acoustic wave equation \eqref{eq:acoustic_initia_value_problem} using the pseudospectral $k$-space method implemented in the k-Wave toolbox \cite{treeby2010} for sensor coverages of $\alpha = 90^\circ, 180^\circ$, and $360^\circ$ (Fig. \ref{fig:setup}) with angular spacing of 1$^\circ$ resulting in 90, 180, and 360 sensor, respectively. The maximum frequency of the simulated ultrasound data was approximately 7 MHz, and the sensor spacing fulfilled the spatial Nyquist criterion for a circular measurement geometry \cite{wang2021}. The speed of sound was $v = 1500$ m/s. The initial pressure was modeled as negligible in the exterior of $\Omega_\mathrm{el}$.
\begin{table}[!tbp]
	\caption{Spatial and temporal discretisations used for data simulation and the acoustic inverse problem. Number of pixels in the reconstruction domain $L$, number of pixels in the acoustic domain $L_\mathrm{ac}$, pixel size $\Delta h$, PML thickness in pixels, number of time points $T$, and time step $\Delta t$.}
	\label{tab:discretisations_acoustic}
	\centering	
	\begin{tabular}{lcccccc}		
		& $L$ & $L_\mathrm{ac}$ & $\Delta h\, (\mu\mathrm{m})$ & PML & $T$ & $\Delta t\, \mathrm{(ns)}$ \\
		\midrule
		Simulation &  & $643^2$ & 17.1 & 16 & 3032 & 3.4 \\
		Inverse    & 9705 & $123^2$ & 90.0 & 12 & 580 & 18.0  \\
		\midrule 
	\end{tabular}
\end{table}

After data simulation, Gaussian noise was added to all datasets. The noise was defined as zero mean and the standard deviation of the noise was chosen as the average of 1\% of the maximum simulated ultrasound amplitudes for all incident electric fields $E_\mathrm{inc}$ using the sensor geometry $\alpha = 360$.

\subsection{Acoustic inverse problem}
\label{subsec:Acoustic inverse problem simulation}
For the acoustic inverse problem, the simulated thermoacoustic ultrasound data was linearly interpolated to a coarser temporal discretisation described in Table \ref{tab:discretisations_acoustic} and the initial pressure distributions were estimated using \eqref{eq:acoustic_posterior_mean}-\eqref{eq:acoustic_posterior_cov} in a circular 5 mm radius domain corresponding to $\Omega_\mathrm{el}$. The initial pressure distributions were estimated for all incident electric fields $E_\mathrm{inc}$ and angular sensor coverages $\alpha = 90^\circ, 180^\circ$, and $360^\circ$. If the initial pressure included negative values, those were set to zero.

\begin{figure}[!tbp]
	\centering
	\includegraphics{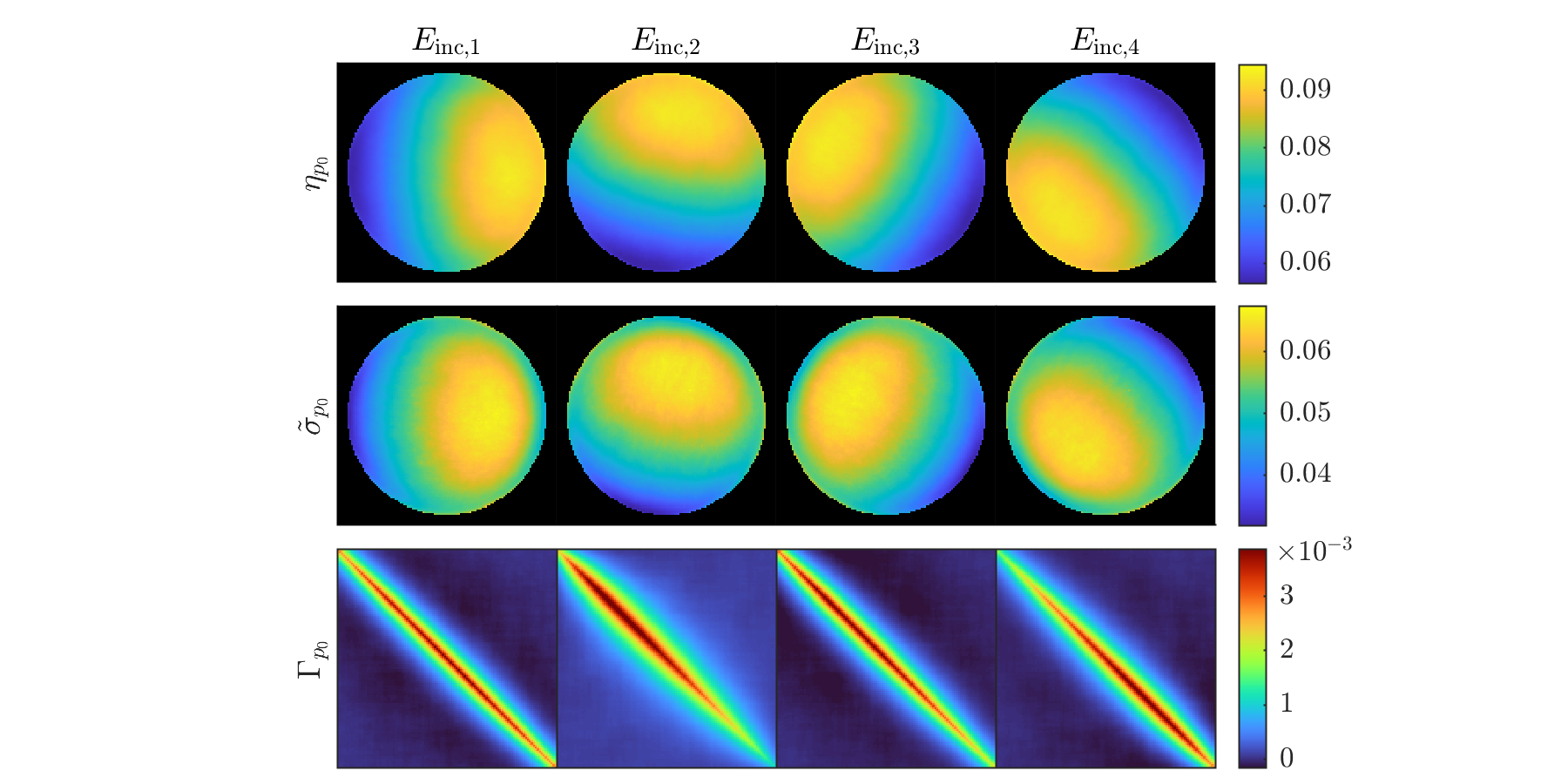}
	\caption{Expected values $\eta_{p_0}$ (first row), standard deviations $\tilde{\sigma}_{p_0}$ (second row), and blocks of the covariance matrices $\Gamma_{p_0}$ (third row) of the prior distribution of the initial pressure in the acoustic inverse problem using the incident electric fields $E_\mathrm{inc,1}$ to $E_\mathrm{inc,4}$ (columns 1-4). For the covariance matrices, a subsection of the covariance matrix corresponding to the pixels illustrated in Fig. \ref{fig:phantoms} is shown.}
	\label{fig:prior}
\end{figure}

The noise was modelled as Gaussian distributed with zero mean. The standard deviation of the noise was modeled as the average of 1\% of the maximum simulated ultrasound amplitudes for all incident electric fields $E_\mathrm{inc}$ using the sensor geometry $\alpha = 360$. The prior distribution \eqref{eq:sample_prior_eta}-\eqref{eq:sample_prior_gamma} was simulated by drawing $N = 50000$ samples $(\hat{\sigma}_n, \hat{\epsilon}_{r,n})_{n = 1,\dots,50000}$ from the Ornstein-Uhlenbeck prior distributions for the dielectric parameters \eqref{eq:dielectric_prior_Gamma}-\eqref{eq:dielectric_prior_Pi}. The samples of the prior distributions for the conductivity and relative permittivity were obtained as $\hat{\sigma}_n = \eta_\sigma + L_{\sigma}^\mathrm{T}\tilde{\epsilon}_n$ and $\hat{\epsilon}_{r,n} = \eta_{\epsilon_r} + L_{\epsilon_r}^\mathrm{T}\tilde{\epsilon}_n$, where $\tilde{\epsilon} \sim \mathcal{N}(0, I)$, and $L_{\sigma} \in \mathbb{R}^{J \times J}$ and $L_{\epsilon_r} \in \mathbb{R}^{J \times J}$ are the Cholesky decompositions of the covariance matrices $\tilde{\sigma}^2_\sigma\Pi$ and $\tilde{\sigma}^2_{\epsilon_r}\Pi$ (Eq. \eqref{eq:dielectric_prior_Gamma}), respectively (Table \ref{tab:prior_parameters}). Initial pressure samples $p_{0,n}$ were then simulated using the FEM in the inverse FE discretization described in Table \ref{tab:discretisations_electromagnetic}, and stored in memory for computation of the prior mean and covariance for  the initial pressure \eqref{eq:sample_prior_eta}-\eqref{eq:sample_prior_gamma}. The expected values $\eta_{p_0 \vert p_t}$, standard deviations $\tilde{\sigma}_{p_0 \vert p_t}$, and blocks of the covariance matrices $\Gamma_{p_0 \vert p_t}$ of the prior distribution for all incident electric fields $E_{\mathrm{inc}}$ are visualized in Fig. \ref{fig:prior}. As it can be seen, the expected values and standard deviation exhibit significant differences depending on the direction of the incident electric field. Furthermore, the covariance matrices are highly non-diagonal indicating a degree of spatial correlation. We note that although constructing the prior model by sampling and evaluating forward solutions can be computationally expensive, it can be done off-line prior to solving the inverse problem and that the same prior model can be used as long as the measurement setup remains the same and the imaged target is described by the prior model.
\begin{table}[!tbp]
	\caption{Expected values $\eta$, standard deviations $\tilde{\sigma}$, and characteristic lengths $\ell$ of the prior distributions for the conductivity $\sigma$ and the relative permittivity $\epsilon_r$.  }
	\label{tab:prior_parameters}
	\centering	
	\begin{tabular}{lccc}		
		& $\eta$ & $\tilde{\sigma}$ & $\ell$ (mm)  \\ 
		\midrule
		Conductivity $\sigma$                 & 0.87 & 0.49 & 1.00\\
		Relative permittivity $\epsilon_r$    & 34.74 & 19.53 & 1.00  \\
		\midrule
	\end{tabular}
\end{table}

\begin{figure}[!tb]
	\centering
	\includegraphics{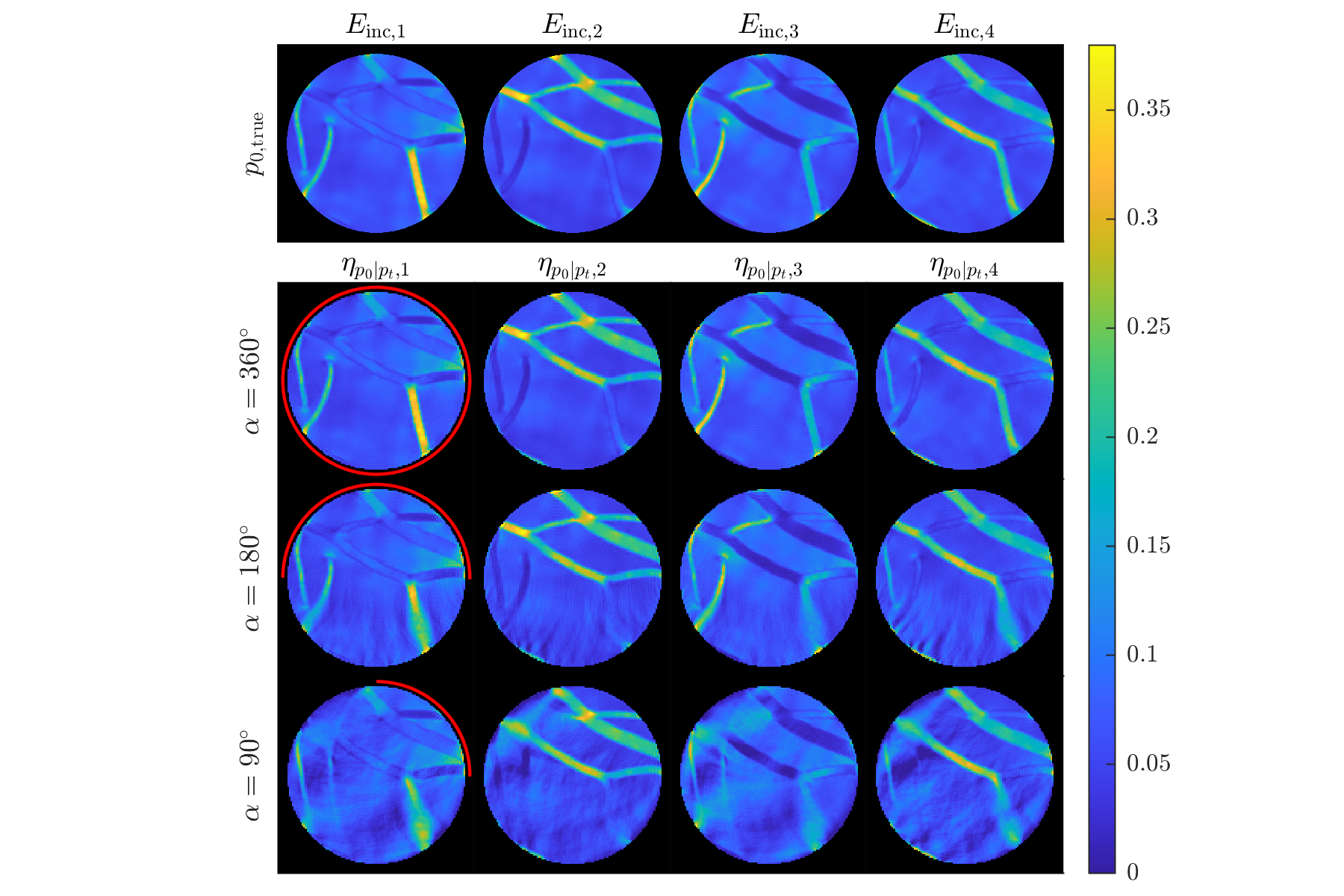}
	\caption{Simulated (true) initial pressure $p_0$, (first row) and estimated initial pressures $\eta_{p_0 \vert p_t}$ (rows 2-4) for all incident electric fields $E_\mathrm{inc}$ (columns 1-4) and sensor geometries $\alpha = 90, 180$, and $360$. The sensor geometries are illustrated using red arcs.}
	\label{fig:acoustic}
\end{figure}

The estimated expected values $\eta_{p_0 \vert p_t}$ of the initial pressure for angular sensor coverages $\alpha = 90^\circ, 180^\circ$, and $360^\circ$ and all incident electric fields $E_\mathrm{inc}$ are shown in Fig. \ref{fig:acoustic}. As it can be seen, amplitudes of the initial pressure vary significantly depending on the vessel orientation with respect to the incident electric field direction. Furthermore, the effect of the limited view sensor geometries can be observed as blurring of the phantom structures outside the sensor coverage.

\subsection{Electrical inverse problem}
\label{subsec:Electrical inverse problem simulation}
\begin{figure}[!tb]
	\centering
	\includegraphics{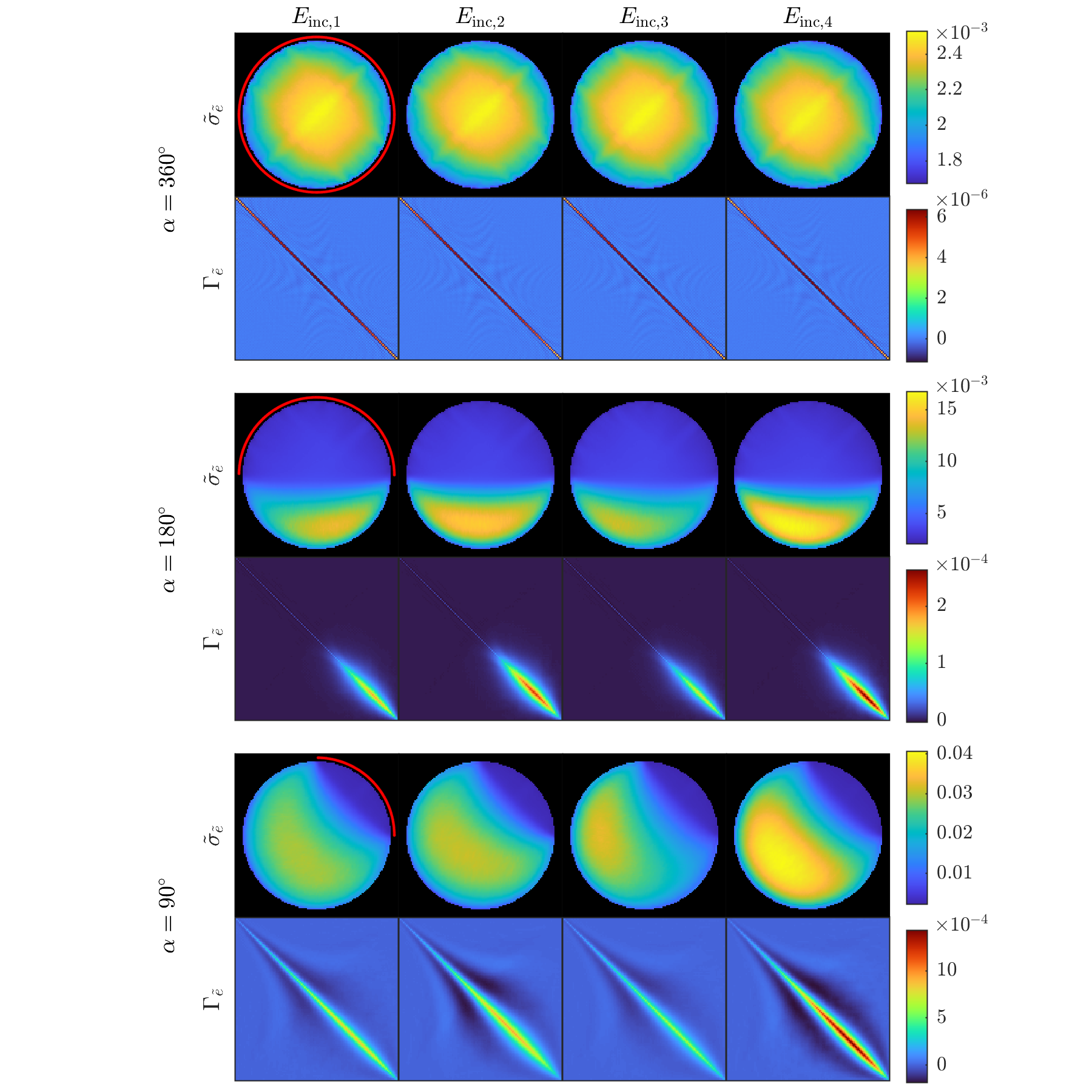}
	\caption{Standard deviations $\sigma_{\tilde{e}}$ and blocks of the covariance matrices $\Gamma_{\tilde{e}}$ of the noise for the electrical inverse problem for incident electric fields $E_\mathrm{inc,1}$ to $E_\mathrm{inc,4}$ (columns 1-4), and sensor geometries $\alpha = 90, 180$, and $360$ (rows 1-6). For the covariance matrices, a subsection of the covariance matrix corresponding to the pixels indicated in Fig. \ref{fig:phantoms} is shown. The sensor geometries are illustrated using red arcs.}
	\label{fig:acoustic_covariance}
\end{figure}

The data for the electrical inverse problem was given by the expected value of the initial pressure distribution (Eq. \eqref{eq:acoustic_posterior_mean}). Furthermore, the noise was given by the covariance of the initial pressure (Eqs. \eqref{eq:acoustic_posterior_cov} and \eqref{eq:electrical_noise_statistics}). The standard deviations $\tilde{\sigma}_{\tilde{e}}$ and blocks of the covariance matrices of the noise $\Gamma_{\tilde{e}}$ for all incident electric fields $E_\mathrm{inc}$ and sensor geometries $\alpha = 90^\circ, 180^\circ$, and $360^\circ$ are illustrated in Fig. \ref{fig:acoustic_covariance}. As it can be seen, the structure of the standard deviation is largely determined by the sensor geometry. The standard deviations are generally largest outside the area covered by the sensor array and vary slightly depending on the direction of the incident electric field. Furthermore, the covariance matrices become increasing non-diagonal as the sensor coverage decreases.

The expected value and standard deviation of the Ornstein-Uhlenbeck prior distribution \eqref{eq:dielectric_prior_Gamma}--\eqref{eq:dielectric_prior_Pi} for the dielectric parameters were chosen as $\eta_\sigma = \frac{1}{2}(\sigma_{\mathrm{max}} + \sigma_{\mathrm{min}})$,  $\eta_{\epsilon_r} = \frac{1}{2}(\epsilon_{r,\mathrm{max}} + \epsilon_{r,\mathrm{min}})$,  $\tilde{\sigma}_\sigma = \frac{1}{3}(\sigma_{\mathrm{max}} - \sigma_{\mathrm{min}})$, and $\tilde{\sigma}_{\epsilon_r} = \frac{1}{3}(\epsilon_{r,\mathrm{max}} - \epsilon_{r,\mathrm{min}}),$ where the subscripts max and min refer to the maximum and minimum values of the phantom. The characteristic length was $\ell = 1\, \mathrm{mm}$. Numerical values of the prior distribution parameters for the conductivity and relative permittivity are shown in Table \ref{tab:prior_parameters}.

The electrical conductivity $\sigma$ and relative permittivity $\epsilon_r$ were estimated by minimizing \eqref{eq:map_estimate}. The dielectric parameters were estimated using data from one to four incident electric fields $E_\mathrm{inc}$ and angular sensor coverages of $\alpha = 90^\circ, 180^\circ$, and $360^\circ$. The minimization problem \eqref{eq:map_estimate} was solved with the Gauss-Newton method \eqref{eq:gauss_newton_iteration}. The minimization problem was constrained to $\sigma \geq 0$ and $\epsilon_r \geq 1$, and the initial guess of the minimization algorithm was set at the expected value of the prior distribution. The minimization problem was iterated for a minimum of 10 iterations. The iterations were stopped if the objective function value increased for three consecutive iterations or maximum number of 25 iterations was reached. 

\begin{figure*}[!p]
	\centering
	\includegraphics[]{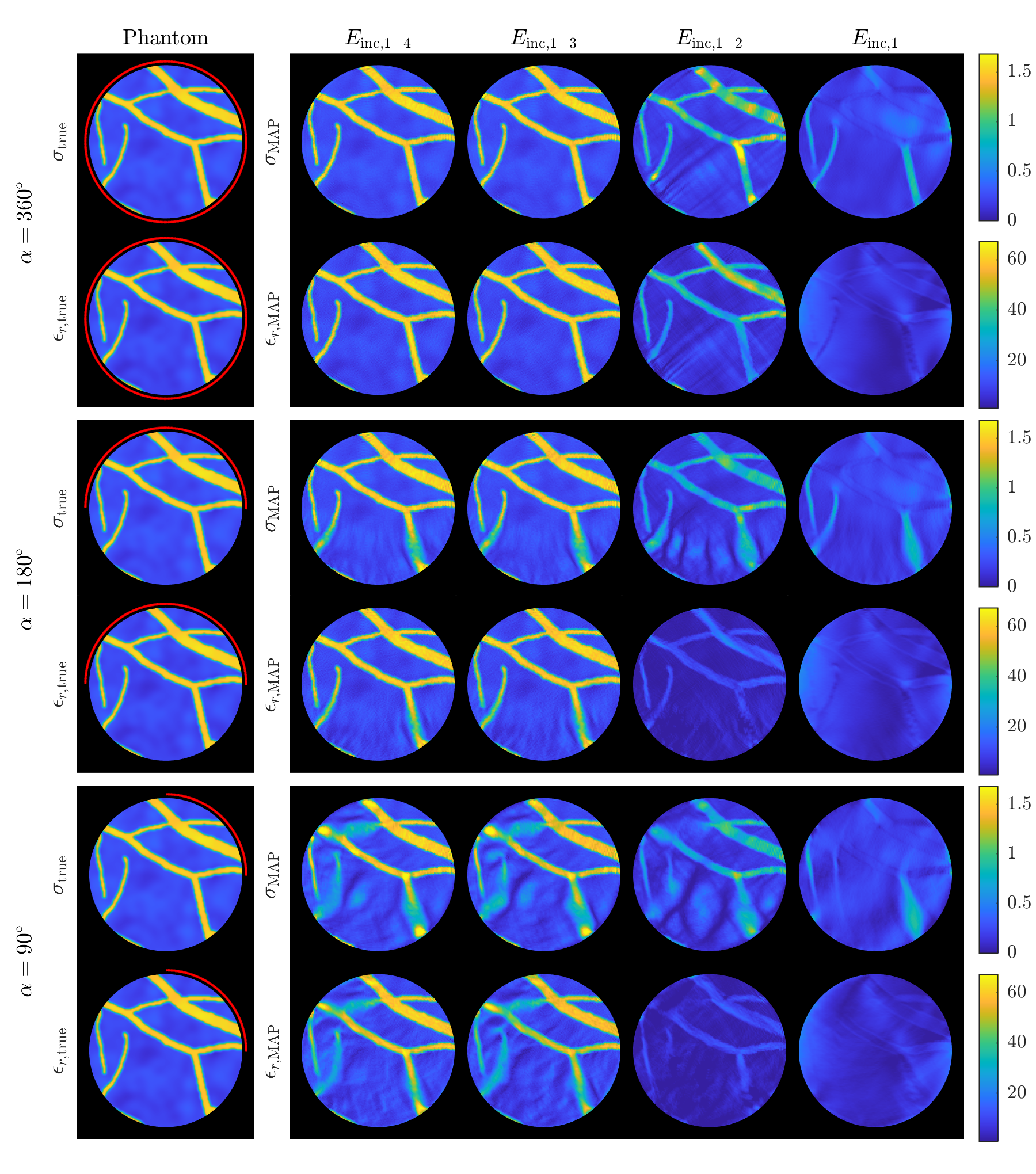}
	\caption{MAP estimates for conductivity $\sigma_\mathrm{MAP}$ (rows 1, 3, and 5) and relative permittivity $\epsilon_{r,\mathrm{MAP}}$ (rows 2, 4, and 6) for one to four incident electric fields $E_\mathrm{inc}$ (columns 2-5) and sensor geometries $\alpha = 360^\circ, 180^\circ$, and $90^\circ$ (rows 1-2, 3-4, and 5-6). The true dielectric parameters $\sigma_\mathrm{true}$ and $\epsilon_{r,\mathrm{true}}$ are shown in the first column. The sensor geometries are shown using red arcs.}
	\label{fig:electrical}
\end{figure*}

\begin{figure*}[!p]
	\centering
	\includegraphics[]{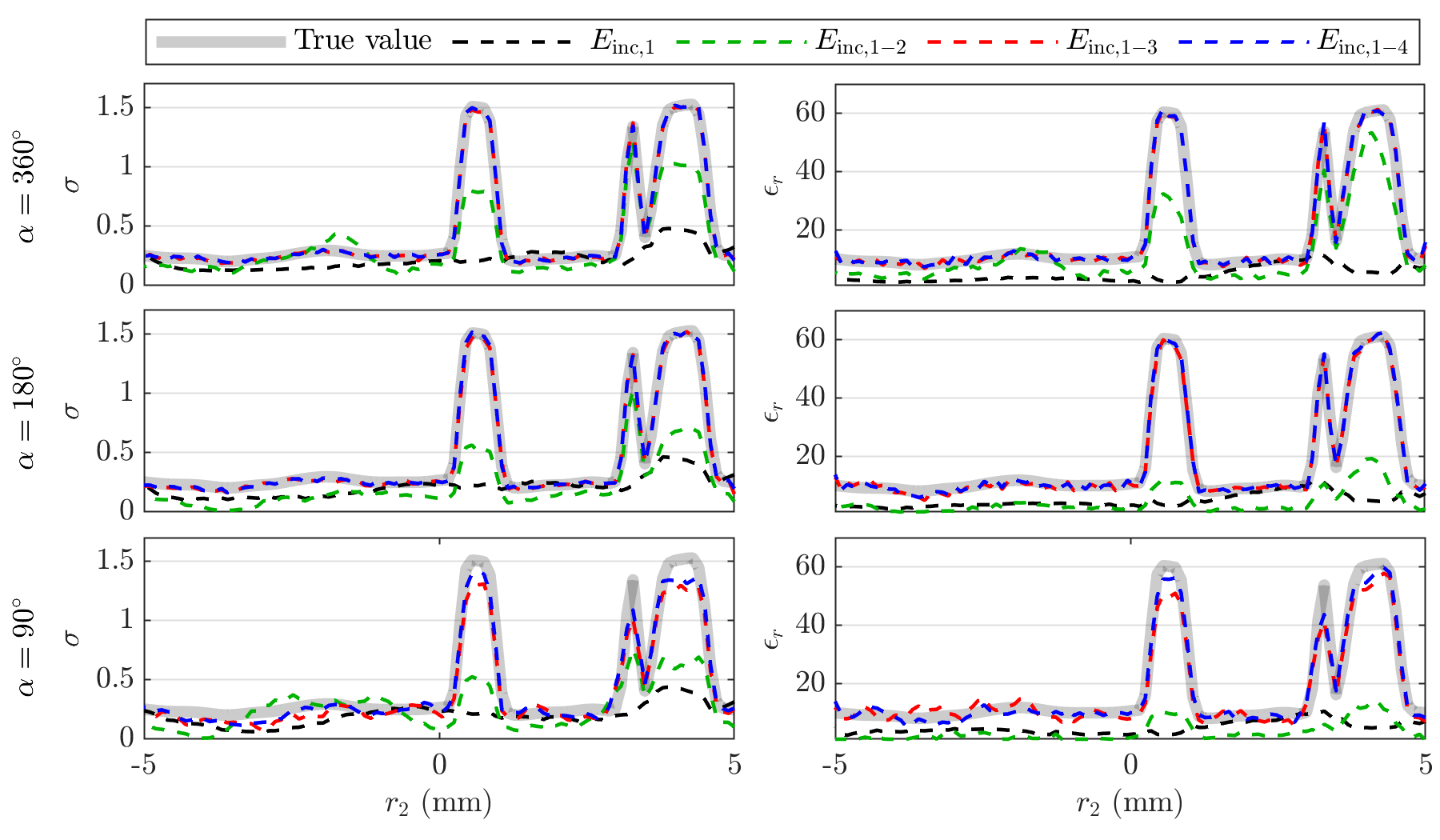}
	\caption{Cross sections of the MAP estimates for conductivity $\sigma_\mathrm{MAP}$ (first column) and relative permittivity $\epsilon_{r,\mathrm{MAP}}$ (second column) using one to four incident electric fields $E_\mathrm{inc}$ for sensor geometries $\alpha = 90^\circ, 180^\circ$, and $360^\circ$ (rows 1-3). The parameters are visualized on a vertical cross section through the domain illustrated in Fig. \ref{fig:phantoms}. }
	\label{fig:electrical_cross_section}
\end{figure*}

The MAP estimates of the conductivity $\sigma_\mathrm{MAP}$ and relative permittivity $\epsilon_{r,\mathrm{MAP}}$ for one to four incident electric field $E_\mathrm{inc}$ and sensor geometries $\alpha = 90^\circ, 180^\circ$, and $360^\circ$ are shown in Fig. \ref{fig:electrical}. Corresponding cross-sectional line plots are shown in Fig. \ref{fig:electrical_cross_section}.

As it can be seen, the number of incident electric fields has a significant effect on the accuracy of the estimated parameters. In the case of one incident electric field, the estimates are highly inaccurate for all sensor geometries. In the case of two incident electric fields, relatively accurate estimated can be obtained when using the full view sensor geometry. Furthermore, in the case of three and four incident electric fields, the estimates are close to the true values for the full view sensor geometry. Furthermore, the effect of the limited ultrasound sensor geometries can be observed as errors and artifacts in the areas where the limited view artifacts are seen in the estimated initial pressures (Fig. \ref{fig:acoustic}). 


\section{Discussion and conclusions}
\label{sec:Discussion}
In this work, an approach for estimating the dielectric parameters from thermoacoustic ultrasound waves in QTAT was presented. The inverse problem was approached in the Bayesian framework and was solved in two stages. The approach was studied using numerical simulations with varying number of incident electric fields, and full and limited view ultrasound sensor geometries. 

The results of this study highlight two major factors influencing the accuracy of the estimated dielectric parameters in QTAT. First, the number of electromagnetic pulses has a significant effect on the estimated dielectric parameters. In previous works on QTAT it has been shown that multiple electromagnetic pulses inducing sufficiently different initial pressures are required for obtaining a unique solution to the electrical inverse problem \cite{bal2014,sahlstrom2025}. In this work, we observed that three electromagnetic pulses from different directions around the target were sufficient to achieve accurate estimates, whereas two pulses resulted in reasonably accurate estimates when using a full view ultrasound sensor geometry. The optimal number of pulses is, however, likely to depend on factors such as the chosen field directions, pulse frequency, polarization, and the properties of the imaged target. Second, the geometry of the ultrasound sensor array has a significant effect on the reconstruction quality. While accurate estimates can be achieved within the region enclosed by the sensors, estimates suffer from artifacts further away from the sensor array. These errors originate from limited view artifacts present in the solution of the acoustic problem that can further be exacerbated when using sparsely positioned sensor arrays. 

To move towards realistic measurement systems, characteristics of different practical measurement systems should be considered. These include the finite size and bandwidth of ultrasound sensors, and sparsely positioned sensor arrays, which may lead to artifacts in the solution of the acoustic inverse problem. Moreover, the use of plane wave electric fields and the assumption of instantaneous electromagnetic energy absorption may not accurately represent practical systems, particularly when the target is positioned near a realistic waveguide or antenna source. 

In this work, the inverse problem of QTAT was studied using 2D simulations. As propagation of acoustic and electromagnetic waves are inherently three dimensional, the proposed approach should be extended to a three dimensional setting where realistic ultrasound sensors and more realistic electromagnetic models, boundary conditions, and electromagnetic sources can be taken into account. This can, however, be computationally expensive, particularly in large computational domains. In these cases, precomputation and storage of the acoustic forward operator may become infeasible necessitating the use of matrix free implementations \cite{tick2018}. Furthermore, computation of the sample based prior distribution can become computationally prohibitive when sufficiently dense discretisations or large computational domains are used. In these cases, alternative sampling techniques \cite{vono2022} and approximative models for the prior covariances and electromagnetic forward model could be considered.

Biological tissues generally exhibit frequency dependent dielectric properties, and the propagation of electromagnetic waves in a complex medium can change significantly with the frequency of the applied electric field. In this work, the inverse problem was studied using an electric field frequency of 1 GHz, which offers a favorable trade-off between penetration depth and electromagnetic contrast. To gain additional insight into the proposed method, future studies should account for the frequency dependent dielectric parameters and investigate the problem using multiple electric field frequencies.

Overall, it can be concluded that the dielectric parameters can be accurately estimated using the proposed approach and that relatively accurate estimates can be obtained even in severely limited view ultrasound sensor geometries as long as sufficiently many electromagnetic pulses are used.

\section*{Acknowledgments}
This work was supported by the European Research Council (ERC) under the European Union’s Horizon 2020 Research and Innovation Program (Grant Agreement No. 101001417-QUANTOM) and the Research Council of Finland (Centre of Excellence in Inverse Modelling and Imaging grant 353086, and Flagship of Advanced Mathematics for Sensing Imaging and Modelling grant 358944), and Orion Science Foundation. The authors would like to thank Professor Amelie Litman and Timo Lähivaara for insightful discussions and suggestions. 



\end{document}